\documentclass[usenatbib]{mnras}
\usepackage[ruled,vlined]{algorithm2e}
\usepackage{subfig}

\SetKwComment{tcp}{\% }{}
\DontPrintSemicolon

\usepackage{newtxtext,newtxmath}

\usepackage[T1]{fontenc}

\DeclareRobustCommand{\VAN}[3]{#2}
\let\VANthebibliography\thebibliography
\def\thebibliography{\DeclareRobustCommand{\VAN}[3]{##3}\VANthebibliography}

\usepackage{graphicx}	
\usepackage{amsmath}	

\title[jFoF: GPU Cluster Finding with Gradient Propagation]{jFoF: GPU Cluster Finding with Gradient Propagation}

\author[B. Horowitz, A. E. Bayer]
{
Benjamin Horowitz$^{1,2,3}$\thanks{E-mail: ben.horowitz@ipmu.jp}, Adrian E. Bayer$^{4,5}$
\\
$^{1}$Kavli IPMU (WPI), UTIAS, The University of Tokyo, Kashiwa, Chiba 277-8583, Japan \\
$^{2}$Center for Data-Driven Discovery, Kavli IPMU (WPI), UTIAS, The University of Tokyo, Kashiwa, Chiba 277-8583, Japan\\
$^{3}$Lawrence Berkeley National Lab, 1 Cyclotron Road, Berkeley, CA 94720, USA\\
$^{4}$Center for Computational Astrophysics, Flatiron Institute, 162 5th Avenue, New York NY 10010, USA\\
$^{5}$Department of Astrophysical Sciences, Peyton Hall, Princeton University, Princeton, NJ 08544, USA\\
}

\date{Accepted XXX. Received YYY; in original form ZZZ}

\pubyear{\the\year{}}

\begin{document}
\label{firstpage}
\pagerange{\pageref{firstpage}--\pageref{lastpage}}
\maketitle

\begin{abstract}

We present \texttt{jFoF}, a fully GPU-native Friends-of-Friends (FoF) halo finder designed for both high-performance simulation analysis and differentiable modeling. Implemented in \texttt{JAX}, \texttt{jFoF} achieves end-to-end acceleration by performing all neighbor searches, label propagation, and group construction directly on GPUs, eliminating costly host–device transfers. We introduce two complementary neighbor-search strategies, a standard $k$-d tree and a novel linked-cell grid, and demonstrate that \texttt{jFoF} attains up to an order-of-magnitude speedup compared to optimized CPU implementations while maintaining consistent halo catalogs. Beyond performance, \texttt{jFoF} enables gradient propagation through discrete halo-finding operations via both frozen-assignment and topological optimization modes. Using a topological optimization approach via a REINFORCE-style estimator, our approach allows smooth optimization of halo connectivity and membership, bridging continuous simulation fields with discrete structure catalogs. These capabilities make \texttt{jFoF} a foundation for differentiable inference, enabling end-to-end, gradient-based optimization of structure formation models within GPU-accelerated astrophysical pipelines. We make our code publicly available at \url{https://github.com/bhorowitz/jFOF/}.

\end{abstract}




\section{Introduction}

Dark matter halos are the fundamental building blocks of cosmic structure \citep{1974ApJ...187..425P}. They provide the gravitational potential wells within which galaxies form, merge, and evolve, linking the small-scale physics of galaxy formation to the large-scale distribution of matter in the Universe. \citep{1972ApJ...176....1G,1993MNRAS.262..627L,2010gfe..book.....M} The abundance \citep{1978MNRAS.182...27D,2001MNRAS.321..372J,2008ApJ...688..709T}, clustering \citep{1986ApJ...304...15B}, and internal properties \citep{1986Natur.322..329Q,1994ApJ...434..402C,1996ApJ...462..563N,1997ApJ...490..493N} of halos encode valuable information about the underlying cosmological model, including the amplitude and shape of the matter power spectrum \citep{2001MNRAS.321....1W}, the growth of structure \citep{2009ApJ...692..217L}, and the nature of dark energy \citep{2012MNRAS.424..993C}. As a result, accurate identification and characterization of halos in cosmological simulations are central to both galaxy modeling \citep{2000MNRAS.318.1144P} and theoretical predictions for large-scale structure observables such as galaxy clustering and weak lensing \citep{2008ApJ...688..709T,2011MNRAS.415.2293K}. 

The Friends-of-Friends (FoF) algorithm \citep{1982ApJ...257..423H} has long been one of the most widely used methods for identifying halos in N-body simulations \citep{1985ApJ...292..371D}. In its classical form, FoF links all pairs of particles separated by less than a specified linking length, effectively tracing isodensity surfaces in the underlying particle distribution. The result is a percolation-based clustering that naturally partitions the simulation volume into disjoint overdense regions. Despite its conceptual simplicity, the FoF method remains remarkably powerful and robust, and has been employed in nearly all major cosmological simulations including Millennium \citep{2009MNRAS.398.1150B,2010MNRAS.406.2267F}, Illustris \citep{2014MNRAS.444.1518V}, and Bolshoi \citep{2016MNRAS.462..893R} projects. Efficient implementation of FoF typically relies on spatial data structures such as hash tables \cite{2018A&C....25..159C} or $k$-d trees \citep{bentley1975multidimensional}, which reduce the naïve cost of pairwise searches. These algorithmic optimizations have made FoF a standard tool for large-scale halo catalogs and merger-tree construction.

A closely related class of group-finding algorithms has been developed for observational datasets, where the goal is to identify galaxy systems or clusters from redshift-space catalogs \citep{1982ApJ...257..423H}. These methods often extend the FoF formalism to account for redshift distortions and selection effects, employing anisotropic linking lengths or probabilistic membership assignment \citep{2005ApJ...630..759M,2006ApJS..167....1B,2016A&A...588A..14T}. Such observational group catalogs have proven essential for connecting galaxies to their host halos through Halo Occupation Distribution (HOD) models and for constraining galaxy evolution via galaxy group clustering statistics \citep{2006ApJ...638L..55Y,2006MNRAS.366....2W}. 

Modern cosmological analysis is increasingly targeted on GPU architectures to exploit massive parallelism in data analysis, particle–mesh and hydrodynamical solvers, but this shift imposes new design constraints: host–device data movement can dominate runtime unless analysis is GPU-native. Spatial trees (octrees, $k$-d trees) have been ported to GPUs \citep{2022arXiv221100120W}, yet their recursive control flow and irregular memory access hinder efficient vectorization; by contrast, friends-of-friends (FoF) grouping maps cleanly onto SIMD-style kernels and is therefore a natural candidate for acceleration. The need for GPU-resident analysis is amplified by end-to-end GPU workflows—e.g. PMWD \citep{2022arXiv221109958L}, GRAM-X \citep{2023CQGra..40t5009S}, and AthenaK\citep{2024arXiv240916053S}—where keeping data on device enables in-situ halo identification, sub-sampling, and real-time statistics without communication overheads. 


The development of GPU-accelerated computation has gone hand in hand with the rise of automatic differentiation in the machine learning community, where AD enables efficient gradient propagation through arbitrarily complex programs. This synergy has inspired the emergence of differentiable physical simulations, where gradients with respect to initial conditions or model parameters can be computed directly through the simulation code. Examples of these tools for simulation pipelines include fast particle mesh dark matter (i.e. FlowPM \citep{2021A&C....3700505M}, PMWD \cite{2022arXiv221109958L}), static mesh cosmological hydrodynamics \citep{2025arXiv250202294H}, as well as for explicit modeling of instrumental systematics (e.g. \citet{2025A&C....5100930W,2025MNRAS.541..687K}). The gradients provided by these codes can be used in gradient-based optimizers or samplers to perform inference (see e.g.~\cite{MCLMC}). These types of tools can be used for a variety of tasks including reconstructing the local universe \citep{2013MNRAS.432..894J,2016ApJ...831..164W} from galaxy catalogs and the $z \sim 2.0$ universe from Ly$\alpha$ forest data \citep{2019TARDIS,2021TARDISII,2022ApJS..263...27H}. However, a major obstacle remains: many key operations in physical pipelines — such as halo finding — are inherently discrete and thus not differentiable. Thus, approximate models based on bias expansions are often employed to enable differentiability.

Recent approaches have begun to \emph{learn} halo catalogs directly from approximate simulations, bypassing explicit group finding. \citet{2024arXiv240911401P} propose a transformer-based conditional generative model that maps particle-mesh (PM) density fields to 3D halo positions and masses, recovering small-scale halo statistics at the few-percent level while noting that conventional N-body and halo-finding steps remain computationally expensive and non-differentiable. In parallel, CHARM \citep{2024arXiv240909124P} uses a multi-stage, auto-regressive normalizing-flow pipeline to synthesize mock catalogs conditioned on PM fields, learning the number, masses, and velocities of halos in stages to match target statistics without running high-resolution N-body. Together, these works demonstrate that simulation-based emulators can reproduce halo populations efficiently; however, they do not provide a physically exact FoF procedure nor a general route to propagate gradients through the discrete halo-identification step itself. It is likely these learned approaches would have significant cosmological dependencies, requiring large training sets to explore extended cosmological models. Related approaches (e.g., \citet{2025MNRAS.542.1403D}) bypass gradients by training residuals to update neural optimizers. This requires accurate supervised data to map residuals, and in its current form form is used for optimization.

Tree-based and linking algorithms like FoF involve discrete selection, thresholding, and graph connectivity, which all yield piecewise-constant outputs without well-defined analytical derivatives. This poses a problem for differentiable pipelines that aim to connect continuous fields (e.g., density or potential) to discrete observables such as halo particle mass or downstream galaxy membership. Analogous challenges appear in the machine learning community, where discrete decisions occur in reinforcement learning \citep{sutton1998reinforcement}, combinatorial optimization \citep{2018arXiv181106128B}, and neural topology search \citep{2021arXiv210315954H}. To address these issues, researchers have developed gradient estimators such as policy gradients \citep{sutton1999policy,2019arXiv190610652M} and Gumbel–Softmax relaxations \citep{2016arXiv161100712M,2016arXiv161101144J}, which enable gradient propagation through discrete selection. 

In recent astrophysical work, similar ideas have been applied to make discrete simulation steps differentiable. For example, DiffHydro \citep{2025arXiv250202294H} and DiffHOD \citep{2024MNRAS.529.2473H} introduce smooth and probabilistic assignment rules that enable gradients to pass through traditionally non-differentiable grouping and sampling steps. Such differentiable discrete assignments open the possibility of gradient-based calibration for halo models and sub-grid physics. The convergence of these GPU-native computation, automatic differentiation, and probabilistic relaxation motivates a new class of analysis tools designed for both performance and differentiability.

In this work, we present \texttt{jFoF}, a GPU-based Friends-of-Friends halo finder designed for astrophysical and cosmological applications. The algorithm operates natively on GPUs, eliminating CPU–GPU data transfer costs and achieving high parallel efficiency. Moreover, \texttt{jFoF} supports multiple gradient approximations to enable its integration into differentiable astrophysical pipelines. We explore both frozen-assignment gradients, where group membership is treated as fixed during backpropagation, and Gumbel–Softmax–inspired relaxations, which allow smooth, differentiable optimization of halo connectivity and topology. Together, these capabilities demonstrate that fully differentiable structure identification is both feasible and practical at scale, paving the way for new approaches to gradient-based inference and simulation-driven model optimization.

\begin{figure}

\centering
\subfloat[$k$-d-tree]{\label{fig:kd}
\centering
\includegraphics[width=0.8\linewidth]{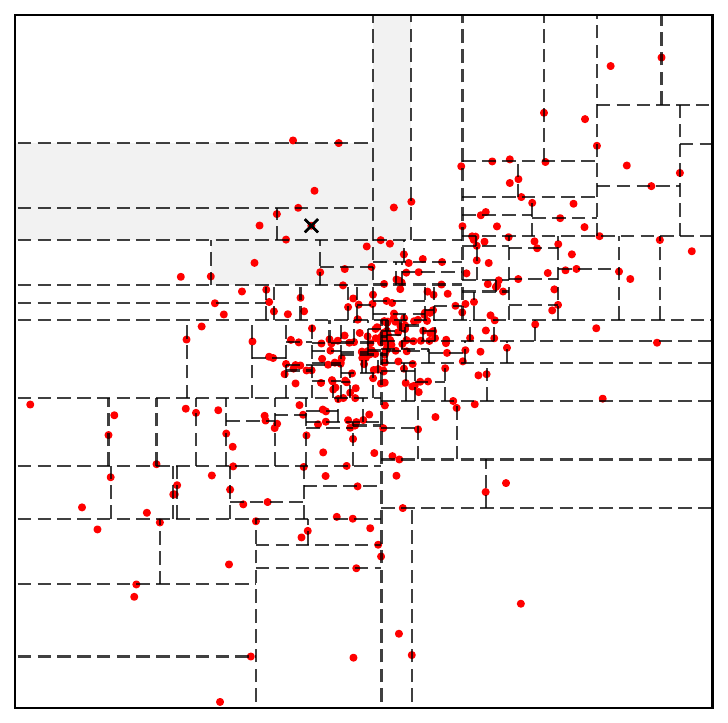}
}
\hfill
\subfloat[linked-cell]{\label{fig:lc}
\centering

\includegraphics[width=0.8\linewidth]{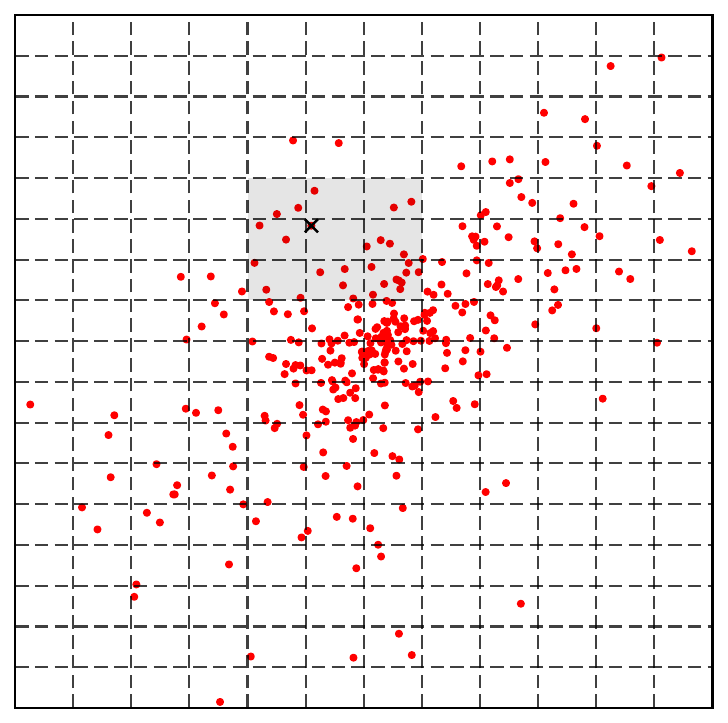}
}

\caption{Schematic illustration of the two spatial data structures used in jFoF for candidate neighbor search. (a) The k–d tree recursively partitions the particle set into axis-aligned hyperrectangles, enabling efficient range queries at arbitrary depths. (b) The linked-cell grid discretizes the simulation volume into fixed cubic cells, limiting neighbor searches to a local stencil of 9 cells (including self) in 2d. The grey boxes denote regions within the search radius ($k_{max}=8$) for a sample query particle.}
    \label{fig:placeholder}
\end{figure}

\section{Halo Finding via Friends of Friends}

There exist a large number of methods for halo/cluster identification that have been developed since early work by \citet{1985ApJ...292..371D} (see \citep{2011MNRAS.415.2293K} for an overview) Along with spherical overdensity, friends-of-friends (FoF) algorithm is the core algorithm behind many commonly used halo finders such as pFoF \cite{pfof} and ROCKSTAR \citep{rockstar}. The fundamental idea of FoF is conceptually simple: given a linking length parameter $b$, two particles are considered “friends” if their separation is less than $b$ times the mean interparticle spacing. Any two friends of friends belong to the same group, resulting in a transitive relation that naturally partitions the particle set into connected components. This approach makes no assumptions about the geometry or density profile of the resulting structures, allowing the algorithm to recover irregular and anisotropic features of the cosmic web. 

To make FoF practical for large datasets typical of modern cosmological simulations, efficient neighbor search and group linking strategies are required. A common acceleration is to use spatial data structures, such as $k$-d trees or cell assignment, to restrict neighbor searches to local regions of space. The $k$-d tree enables hierarchical partitioning of the particle distribution, allowing rapid querying of neighbors within the linking length, while linked cell assignments offer memory-efficient access patterns particularly well-suited to parallelization. We show these methods schematically in Figure \ref{fig:placeholder}.Utilizing one of these two methods (or a combination) can significantly reduce the computational cost of the FoF algorithm, transforming an $\mathcal{O}(n^2)$
 problem into something closer to $\mathcal{O}(n\log n)$. 
 
 Below we briefly describe the common procedure for both of our implementations ($k$-d-tree or a linked cell list);
\begin{enumerate}
    \item \textbf{Input normalization and initialization.} \\
    The positions of all particles are first standardized into a consistent three-dimensional representation. Each particle is assigned a unique initial label, typically its own index, which serves as an identifier that will later propagate across connected regions of the particle distribution during halo construction.

    \item \textbf{Computation of the linking length.} \\
    The dimensionless linking parameter \( b \) is converted into a physical length scale,
\begin{equation}
    l_{\mathrm{link}} = b \left(\frac{V}{N}\right)^{1/3}\,
\end{equation}       
    where \( \bar{l} = (V/N)^{1/3} \) is the mean interparticle spacing derived from the simulation volume \( V \) and total particle number \( N \). This defines the threshold distance below which two particles are considered ``friends.''

    \item \textbf{Identification of candidate neighbors.} \\
    Each particle identifies nearby particles that could potentially fall within the linking length. Depending on the implementation, this may involve:
    \begin{itemize}
        \item constructing a hierarchical data structure such as a \textit{k--d tree} to enable efficient nearest-neighbor queries; or
        \item subdividing the simulation volume into a regular grid (linked-cell list) and searching only the neighboring cells.
    \end{itemize}
    The result is a compact set of candidate neighbors for each particle, bounded by a fixed maximum number or cell occupancy limit.

    \item \textbf{Application of the linking criterion.} \\
    For each candidate pair, the Euclidean (or periodic minimum-image) distance is computed and compared to \( l_{\mathrm{link}} \). Pairs separated by less than this threshold are declared ``friends'' and are connected by an undirected edge in the implicit FoF graph.

    \item \textbf{Group construction via label propagation.} \\
    Once the connectivity structure is established, groups are identified as the connected components of the graph. Instead of explicitly building adjacency lists or performing recursive searches, the algorithm employs an iterative label-propagation scheme: in each iteration, every particle adopts the minimum label among itself and its linked neighbors. Repeated sweeps cause the labels to converge across connected regions, ensuring that all particles in a given halo share a common label.

    \item \textbf{Group compaction and output.} \\
    After convergence, the unique set of final labels defines the distinct FoF groups. Labels are compacted into a contiguous integer range (e.g. \( 0 \ldots N_{\mathrm{halo}}-1 \)), and the number of particles per group is tallied. The resulting catalog provides per-particle group identifiers, group sizes, and the total halo count, forming the basis for subsequent analysis of halo properties.
\end{enumerate}

To test our various implementations, we have generated cosmological simulation boxes via the \texttt{fastPM} particle mesh algorithm \citep{fastPM}. We use a fixed size box with sidelength 500 $h^{-1}$ Mpc, and vary the particle number between $64^3$ and $512^3$, using ten steps with $B=2$.

\begin{figure}
    \centering
\includegraphics[width=0.99\linewidth]{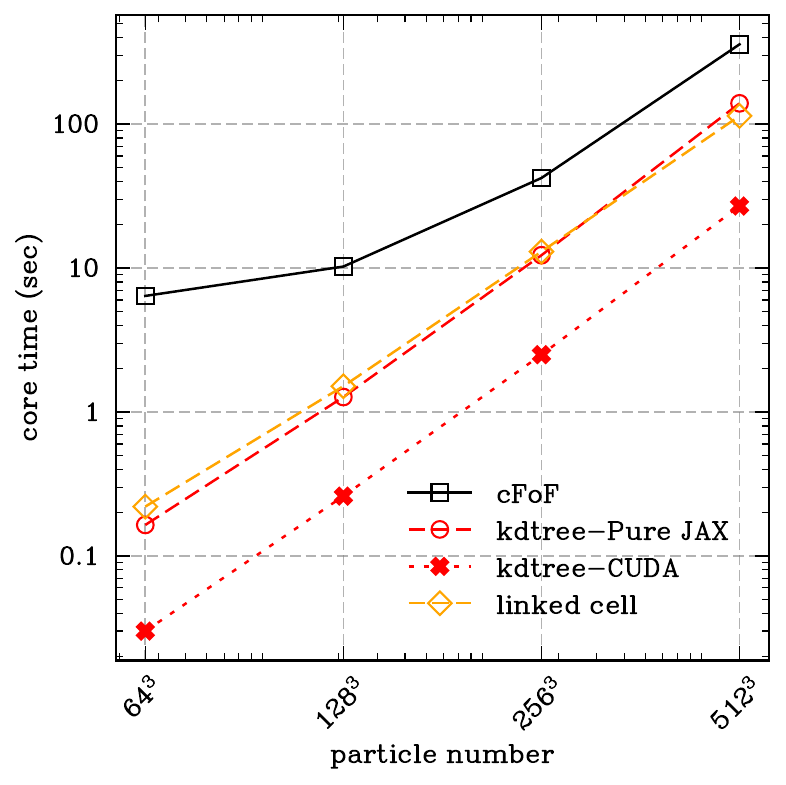}
    \caption{We show the time scalings for various implementations of FoF halo finding on the same dark matter particle mesh simulations generated from \texttt{fastpm} in a $500$ $h^{-1}$ Mpc box. The three \texttt{jFoF} implementations outperform the \texttt{c}-based FoF implementation on a CPU - GPU core hour standpoint by up to an order of magnitude at large particle number.}
    \label{fig:timing}
\end{figure}

We benchmark our implementations against the FoF algorithm included in \texttt{fastpm}, which to avoid ambiguity with its particle mesh routine we will call \texttt{cFoF} (as it is written in C). \texttt{cFoF} provides a parallel implementation of the friends-of-friends algorithm designed to operate efficiently on distributed-memory systems. Its design follows the classic FoF framework but employs \texttt{MPI} (Message Passing Interface) to partition the simulation volume across multiple processes, each responsible for a subdomain of particles. Particles near subdomain boundaries are exchanged between neighboring ranks to ensure that all possible links within the global linking length are correctly identified. Group labels are first determined locally, and a subsequent communication phase merges groups that span multiple domains through iterative label reconciliation across MPI ranks. This strategy allows the algorithm to scale effectively to very large cosmological datasets, maintaining computational efficiency and memory balance while preserving the exact group connectivity that would be obtained in a serial run. We compare the speed of \texttt{cFoF} to various \texttt{jFoF} methods in Figure \ref{fig:timing}. Note we use 128 cores for \texttt{cFoF}.


\subsection{$k$-dtree}
\label{sec:kd}

The \textit{k--d tree} (short for ``$k$-dimensional tree'') is a hierarchical spatial data structure designed to organize points in a $k$-dimensional space for efficient nearest-neighbour and range queries \citep{bentley1975multidimensional}. The structure recursively partitions the space into axis-aligned hyperrectangles by successively splitting the dataset along one coordinate axis at each tree level. Each node stores a subset of the particles and a splitting plane, such that traversal of the tree rapidly excludes large regions of space that cannot contain nearby neighbours. In the context of friends-of-friends halo finding, the k--d tree enables each particle to identify a small subset of candidate neighbours within the linking length without performing an exhaustive pairwise search. Querying the tree typically scales as $\mathcal{O}(\log N)$ per particle, offering a substantial improvement over the brute force $\mathcal{O}(N^2)$ approach. 

Additional thought must be given to GPU-based implementations of the k--d tree algorithm. Unlike a conventional k--d tree built recursively on the CPU, the GPU version must avoid recursion, pointer indirection, and dynamically allocated structures. Instead, it operates on flat arrays with fixed shapes and predictable memory access patterns, making it well-suited for accelerator execution and compatible with \texttt{jax}'s compilation model. We use the \texttt{jaxkd} \footnote{\url{https://github.com/dodgebc/jaxkd/tree/main}} implementation of the tree algorithm, which is based loosely on the \texttt{cudaKDtree} implementation \citep{2022arXiv221100120W}. The \texttt{jaxkd} module provides both a pure \texttt{jax} implementation and a matching cuda-compiled version. \texttt{jFoF} can easily switch between both versions of the code, and we compare the performance against the baseline \texttt{cFoF} and the linked cell implementation (see Sec. \ref{sec:lc}) in Figure \ref{fig:timing}.


We tested both methods also for accuracy in addition to speed. We find near complete agreement between the \texttt{cFoF} and \texttt{jFoF} ktree implementations, with some small differences at the very low mass end (i.e. $\leq 40$ particles) where floating point error could affect the exact halo membership properties. Similarly, in terms of position we find agreement across a range of scales. We test this by painting the halo positions onto a mesh of $500$ sidelength, and calculate the cross correlation coefficient defined as 
\begin{equation}
    r_{AB} = \frac{P_{AB}}{\sqrt{P_{AA} P_{BB}}},
\end{equation}
where $P_X$ is the power spectra of the field and $P_{AB}$ is the cross correlation. We find comparing either \texttt{kdtree} implementation to \texttt{cFoF}, $r_c>0.995$ across all $k$.

\begin{algorithm}[]
\caption{Friends-of-Friends via $k$-d tree (Non-Periodic)}
\label{alg:fof-kdtree-open-knn}
\SetKwInOut{Input}{Input}\SetKwInOut{Output}{Output}
\Input{
  Particle positions $X \in \mathbb{R}^{N\times 3}$; \\
  linking length $b$; nearest-neighbor cap $k_{\max}$; \\
  number of label-relaxation iterations $T$.
}
\Output{
  Cluster labels $\ell \in \{0,\dots,N\!-\!1\}^N$; compact labels $\tilde\ell$; per-cluster sizes.
}
\BlankLine

\textbf{Build $k$-d tree:} Construct a $k$-d-tree $\mathcal{T}$ on the points $X$.\;

\textbf{Neighbor retrieval (per particle):}\;
\For{$i \gets 0$ \KwTo $N-1$}{
  Query the $k_{\max}$ nearest neighbors of $X_i$ from $\mathcal{T}$ (including/excluding $i$ as provided).\;
  Let $\widehat{\mathsf{nbr}}_i$ be the returned indices; remove $i$ if present.\;
  \tcp{Filter by the FoF linking length}
  $\mathsf{nbr}_i \gets \{\, j \in \widehat{\mathsf{nbr}}_i ~:~ \|X_i - X_j\| \le b \,\}$.\;
}

\BlankLine
\textbf{Label propagation (FoF union):}\;
Initialize $\ell_i \gets i$ for all $i$.\;
\For{$t \gets 1$ \KwTo $T$}{
  \For{$i \gets 0$ \KwTo $N-1$}{
    \ForEach{$j \in \mathsf{nbr}_i$}{
      \If{$\|X_i - X_j\| \le b$}{
        $\ell_i \gets \min(\ell_i,\ell_j)$;\quad $\ell_j \gets \min(\ell_j,\ell_i)$\;
      }
    }
  }
}
\BlankLine

\textbf{Compaction:} Map distinct labels in $\ell$ to consecutive integers $\tilde\ell\in\{0,\dots,K'-1\}$ and count occurrences per $\tilde\ell$ to obtain cluster sizes.\;

\end{algorithm}

While this method provides excellent performance for irregular or non-periodic point distributions, it must be adapted when strict periodic boundary conditions are required, as the tree’s axis-aligned splits do not naturally wrap across box boundaries. A significant change to the spatial geometry or introduction of ghost particles would be required to include this topology. While these could be implemented, it would come at some performance cost. In addition it would be unnecessary in the context of realistic field level inference tasks where boundaries would be padded anyway to prevent nonphysical spatial wrapping. We include this wrapping naturally in the linked cell implementation below.

\subsubsection{Sub-Halo Finding}
It is straightforward to extend this formalism to the identification of subhalos using a six-dimensional Friends-of-Friends (FoF) algorithm that operates in particle phase space. Subhalos are critical components of modern abundance-matching approaches to galaxy formation and evolution \citep{2016MNRAS.459.3040G,2019MNRAS.488.3143B}, as they represent the bound structures within larger parent halos that host satellite galaxies. In contrast to standard halo finding, subhalo identification is highly sensitive to the precise algorithm employed, the resolution and force accuracy of the simulation, and the scientific goals of the analysis \citep{2012MNRAS.423.1200O}. Because these details can strongly influence the recovered subhalo mass function and spatial distribution, we do not present a detailed quantitative comparison in this work, focusing instead on illustrating a general, GPU-compatible implementation.

We provide a simplified example inspired by the widely used \texttt{ROCKSTAR} algorithm \citep{rockstar}, designed to demonstrate how a kd tree-based FoF approach can be extended to six-dimensional phase space. We first apply our $k$-d tree FoF method, as described in Sec.~\ref{sec:kd}, to identify primary halos based on spatial proximity. For each identified halo, we then collect the particles associated with it and rerun a second FoF pass that accounts for both particle position, $x_i$, and velocity, $v_i$, treating each as a component of a six-dimensional vector. This allows us to capture substructure that is dynamically coherent but spatially embedded within a larger halo. The phase-space distance metric used to determine particle linking is defined as:
\begin{equation}
d(i,j) = \left( \frac{(\vec{x}_i - \vec{x}_j)^2}{\sigma_x} + \frac{(\vec{v}_i - \vec{v}_j)^2}{\sigma_v} \right)^{1/2}
\end{equation}
where $\sigma_x$ and $\sigma_v$ are normalization factors that set the relative weighting between spatial and velocity separations.

\begin{figure}
    \centering
    \includegraphics[trim={0cm 1cm 0 0},clip,width=0.93\linewidth]{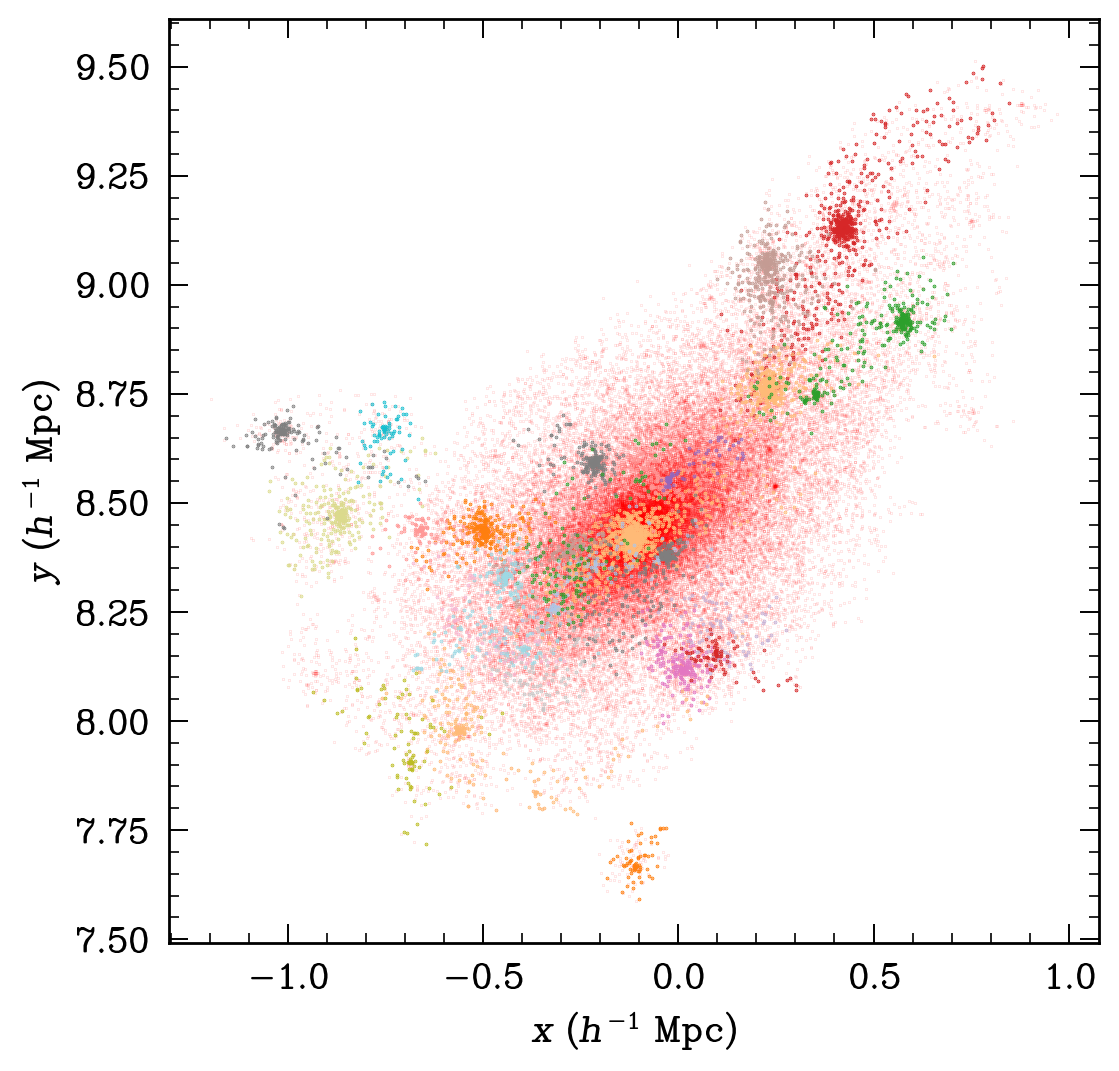}
    \includegraphics[trim={0cm 1cm 0 0},clip,width=0.93\linewidth,]{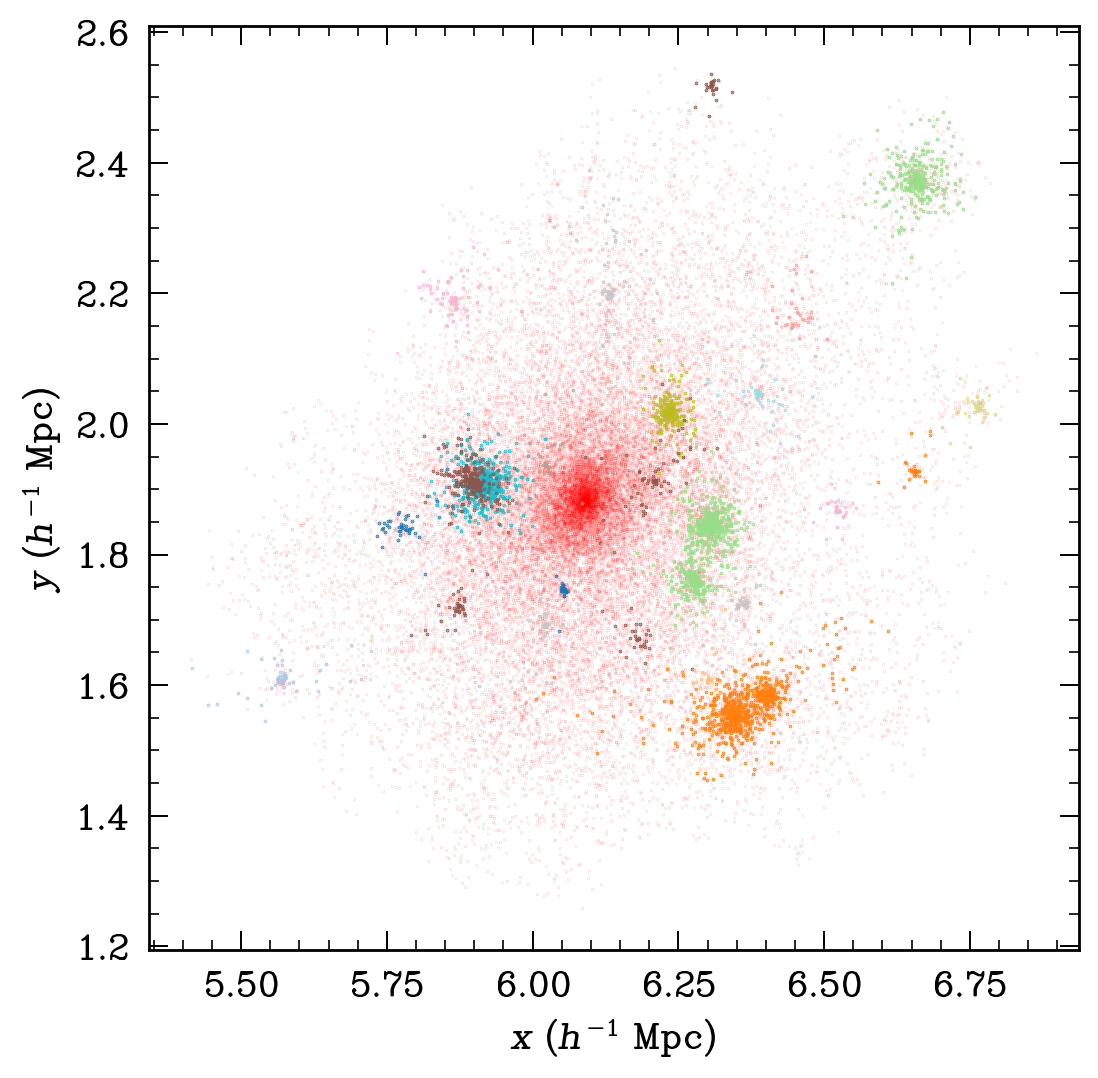}

    \includegraphics[width=0.93\linewidth]{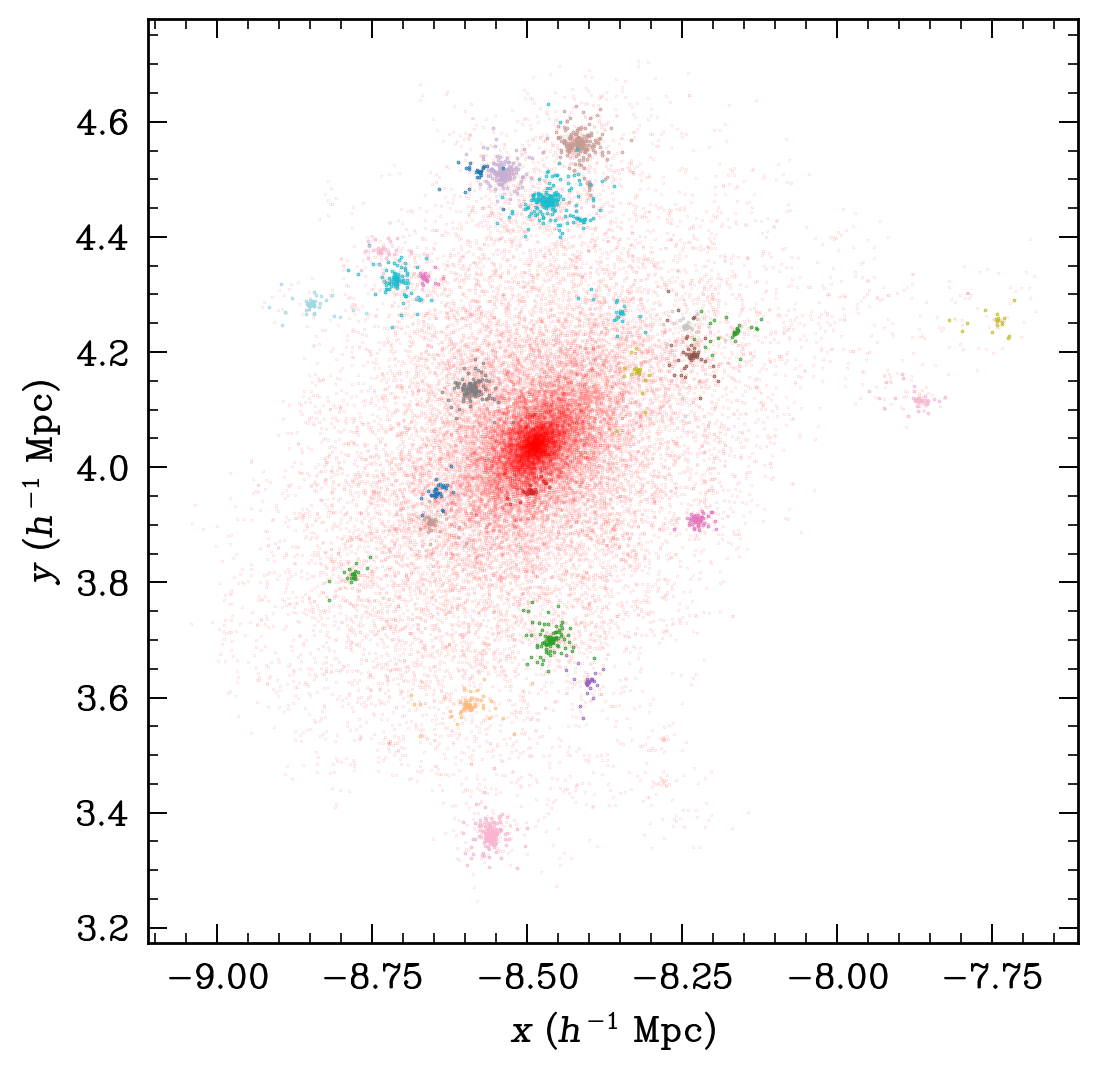}

    \caption{A visualization of halos and sub-halos found in an example CAMELS simulation. We plot the three most massive halos in the volume, and color the 30 largest sub-halos in each with random colors, over-plotted on the total particle distribution. We find our 6d implementation is able to identify merging and merged (sub)structures in phase space.}
    \label{fig:subhalo}
\end{figure}

To demonstrate this implementation, we apply the method to halos identified in the CAMELS simulations \citep{2020arXiv201000619V}, which employ \texttt{Gadget-III} \citep{2005MNRAS.364.1105S} for dark matter evolution. Unlike the \texttt{fastPM} simulations used in earlier sections, \texttt{Gadget-III} properly evolves phase-space information even in highly nonlinear regions during halo mergers. The CAMELS boxes contain $256^3$ particles within a $25~h^{-1}$~Mpc cube, resolving internal halo structures well into the small-scale regime. Figure~\ref{fig:subhalo} shows an example of the largest halos and their associated subhalos identified using this procedure. We find good qualitative agreement with expected substructure morphology, demonstrating that our $k$-d tree-based implementation can recover physically plausible subhalos. The implications of this subhalo-finding method, particularly for constructing merger trees and linking halos across simulation snapshots, will be explored in detail in future work.

\subsection{Linked-Cell Grid Precomputed Neighbors: \texttt{jFoF-LC}}
\label{sec:lc}

In addition to the generic Friends-of-Friends (FoF) $k$-d tree–based algorithms, we implement a high-performance linked-cell variant. This method is similar in spirit to the spatial hashing method introduced for CPU in \cite{2018A&C....25..159C,2018ascl.soft08005C}. The simulation volume is decomposed into a uniform grid of cubic cells, each with a side length comparable to the linking length $b$. This discretization ensures that each particle only needs to test potential links against particles within its own cell and the surrounding 26 neighbors, reducing the neighbor-search complexity from $\mathcal{O}(N^2)$ to $\mathcal{O}(N)$. Periodic boundary conditions are enforced through minimum-image wrapping, maintaining exact equivalence with the canonical FoF connectivity criterion.

Each particle is assigned a linear cell index based on its position modulo the box size $L$. Using this cell map, the algorithm gathers candidate neighbors from the $3^3$ adjacent cells and filters them according to the standard distance threshold $|\mathbf{r}_i - \mathbf{r}_j| \le b$. The implementation supports two distinct execution modes. In precompute mode, a fixed-size neighbor list for all particles is constructed once, enabling rapid label propagation with minimal recomputation. This is directly comparable to the $k$-d tree implementation above. In streaming mode, candidate lists are generated dynamically in small batches, eliminating the need to store the full neighbor table and thus drastically reducing memory consumption. Both modes rely on iterative label minimization (equivalent to a distributed union–find procedure) to propagate group identifiers across all connected particles. For the purpose of this paper, we focus on the precomputed mode which has more optimal performance for relatively small boxes ($\leq 512^3$) used in field level inference works.

The algorithm is written in \texttt{JAX}, enabling vectorized computation and efficient GPU or TPU execution. To preserve numerical stability and deterministic outputs, all operations—cell mapping, sorting, and neighbor filtering—are expressed in terms of pure tensor primitives without control-flow divergence (i.e. nondeterministic execution order due to for loop-type operations). The periodic label propagation step is carried out using fixed-shape memory-safe batches, ensuring scalability to simulations with $N \gtrsim 10^9$ particles.

After convergence, the algorithm remaps the raw particle labels to a compact, contiguous sequence of halo identifiers and accumulates particle counts per halo via a parallel histogram. The resulting catalog is fully equivalent to the standard FoF output, preserving all physical and topological properties of the identified haloes. The linked-cell formulation offers a tunable balance between computational throughput and memory footprint while retaining astrophysical fidelity.

As with the \texttt{kdtree} implementation, we benchmark our method against \texttt{cFoF} similar convergence as discussed in Sec. \ref{sec:kd}. We show the timing performance in Figure \ref{fig:timing}, finding comparable performance to the \texttt{kdtree} pure \texttt{jax} implementation. 

\begin{algorithm}[]
\caption{Periodic FoF with Precomputed Neighbor Lists (Linked-Cell Grid)}
\label{alg:fof-precompute}
\SetKwInOut{Input}{Input}\SetKwInOut{Output}{Output}
\Input{
  Particle positions $X \in \mathbb{R}^{N\times 3}$; box lengths $L\in\mathbb{R}^3$; \\
  linking length $b$; cell cap $\texttt{max\_per\_cell}$; neighbor cap $k_{\max}$; iterations $T$.
}
\Output{
  Cluster labels $\ell \in \{0,\dots,N\!-\!1\}^N$.
}
\BlankLine
\textbf{Wrap positions:} $X \gets X \bmod L$
\textbf{Grid setup:} $n \gets \max(1, \lceil L/b\rceil)$ per axis; cell size $= L / n$\;
\textbf{Cell indexing:} For each $i$, compute cell triple $(i_x,i_y,i_z)$ and linear id
$\,c_i = i_x + n_x(i_y + n_y\, i_z)$.\;
\textbf{Sort by cell:} Form permutation $p$ such that $c_{p(0)} \le \cdots \le c_{p(N-1)}$.\;
\textbf{Sorted ids:} $\hat c_j \gets c_{p(j)}$ for $j=0,\dots,N-1$.\;
\BlankLine
\textbf{Neighbor candidates (per particle):}\;
\For{$i \gets 0$ \KwTo $N-1$}{
  Determine the 27 neighbor cell ids $\mathcal{N}(c_i)$ via $(\cdot \bmod n)$ wrapping.\;
  \tcp{Lookup the range of particles in each neighbor cell using binary search}
  \ForEach{$\gamma \in \mathcal{N}(c_i)$}{
    $s \gets \text{lower\_bound}(\hat c,\ \gamma)$;~~
    $e \gets \text{upper\_bound}(\hat c,\ \gamma)$\;
    Take up to $\texttt{max\_per\_cell}$ indices from $p[s:e]$ and append to $C_i$ (pad with $-1$)\;
  }
  $C_i$ now has fixed length $K = 27\times \texttt{max\_per\_cell}$.\;
  \tcp{Filter by periodic minimum-image distance and linking length}
  Build boolean mask $m_{ij}$ for $j\in[1..K]$ indicating $C_i[j]\ge 0$, $C_i[j]\neq i$, and
  $\|\Delta_{ij}\| \le b$, where $\Delta_{ij}$ uses minimum-image convention.\;
  \tcp{Stable keep of up to $k_{\max}$ neighbors}
  Let $\mathsf{nbr}_i$ be the first $k_{\max}$ indices of $C_i$ with $m_{ij}=1$, padded with $-1$.\;
}
\BlankLine
\textbf{Label propagation (FoF):}\;
Initialize $\ell_i \gets i$.\;
\For{$t \gets 1$ \KwTo $T$}{
  For all edges $(i\leftrightarrow j)$ where $j\in \mathsf{nbr}_i$ and $\|\Delta_{ij}\|\le b$,
  update $\ell_i \gets \min(\ell_i,\ell_j)$ and $\ell_j \gets \min(\ell_j,\ell_i)$.\;
}
\BlankLine
\textbf{Compaction:} Map distinct labels in $\ell$ to consecutive integers $\tilde\ell\in\{0,\dots,K'-1\}$ and count occurrences per $\tilde\ell$ to obtain cluster sizes.\;
\end{algorithm}



\subsection{Parallelization}

We also provide a parallel implementation of Friends-of-Friends (FoF) by decomposing the periodic domain into one-dimensional slabs along a box axis and augmenting each slab with a ghost layer of thickness at least the linking length ($b$). Either a $k$-d tree or a linked–cell grid can be used as the intra–slab neighborhood generator: the slab method is agnostic to that choice. Because both kernels can be organized with fixed shapes (bounded neighbors, fixed batches), the per–slab computation maps cleanly to accelerators and scales with device count; only the non–ghost (center) region of each slab contributes outputs.

Global consistency is achieved by a lightweight fusion step that operates on component labels rather than raw particles. After the per–slab solves, we examine the $\pm b$ overlap bands and, for any particle present in multiple slabs’ working sets, record one equivalence between the two local component labels it belongs to. A single disjoint–set (union–find) pass computes the transitive closure of these equivalences, yielding a mapping from \((\mathrm{slab},\mathrm{local\ label})\) to a deterministic global component identifier (e.g. the minimum global particle ID in the merged set). Relabeling the center particles of each slab via this map produces a globally consistent FoF catalog identical to a monolithic run. Correctness follows because any FoF path that crosses a slab boundary must traverse the overlap and thus induces at least one recorded equivalence.

\begin{table*}
\centering
\small
\renewcommand{\arraystretch}{1.2}
\label{tab:difftable}
\begin{tabular}{p{2.4cm} p{4.5cm} p{4.5cm} p{4.5cm}}
\hline
\textbf{Mode} & \textbf{Description} & \textbf{Differentiable Quantities} & \textbf{Limitations} \\
\hline
Frozen Assignment &
Fixed halo memberships; gradients flow only through particle positions. &
Halo position &
No mass or topology gradients. \\

Decorated Frozen &
Adds analytic “decorations” (e.g.\ particle variance) as property proxies. &
Halo position or property proxies.&
Depends on proxy calibration; still discrete topology. \\

Topological &
Probabilistic linking with stochastic gradients via the REINFORCE estimator. &
Halo position, mass, number, etc. &
High variance; requires tuning of $\alpha$ and sampling. \\
\hline
\end{tabular}
\centering
\caption{Summary of differentiation modes implemented in \texttt{jFoF}.}
\end{table*}
In practice, both neighborhood structures are valid within this framework, but their performance characteristics differ. The slabbed algorithm accommodates either choice (or even a hybrid, e.g. grid inside slabs with $k$-d tree just in boundary bands). Memory is bounded per slab by the neighbor cap and batch sizes, periodic boundaries are handled by wrapping slab edges and overlaps, and load imbalance can be mitigated by non–uniform slab widths or by switching to two–dimensional bricks. The performance characteristics of this approach depend on the choice of segmentation of the original volume and the underlying physics; slab-wise splitting is quite suboptimal. We were able to run $1024^3$ fastPM simulations through this algorithm, but with similar performance time as the CPU based code due to the additional cost of union operations; 3218.8 CPU-seconds for \texttt{cFoF} vs. 4038.0 GPU-seconds for \texttt{jFoF}. 

In practice, since the end goal is coupling \texttt{jFoF} with a GPU-accelerated simulation pipeline, we can follow the particle-division strategy used in the simulator to both minimize GPU-GPU transfer and union-operations. While there are such parallel simulators currently in advanced development (e.g. see \texttt{jaxDecomp} \texttt{jaxpm} \footnote{\url{https://github.com/DifferentiableUniverseInitiative/JaxPM/tree/main}}), there are none currently published in which to benchmark this technique and we leave full exploration to future work.

\section{Gradient Propagation}
While clustering algorithms are analytically non-differentiable, their gradient information can be highly valuable for optimization and inference in astrophysics and cosmology. In this section, we explore several practical strategies for propagating gradients through our halo-finding routine.

\begin{figure*}
    \centering
    \includegraphics[width=0.95\linewidth]{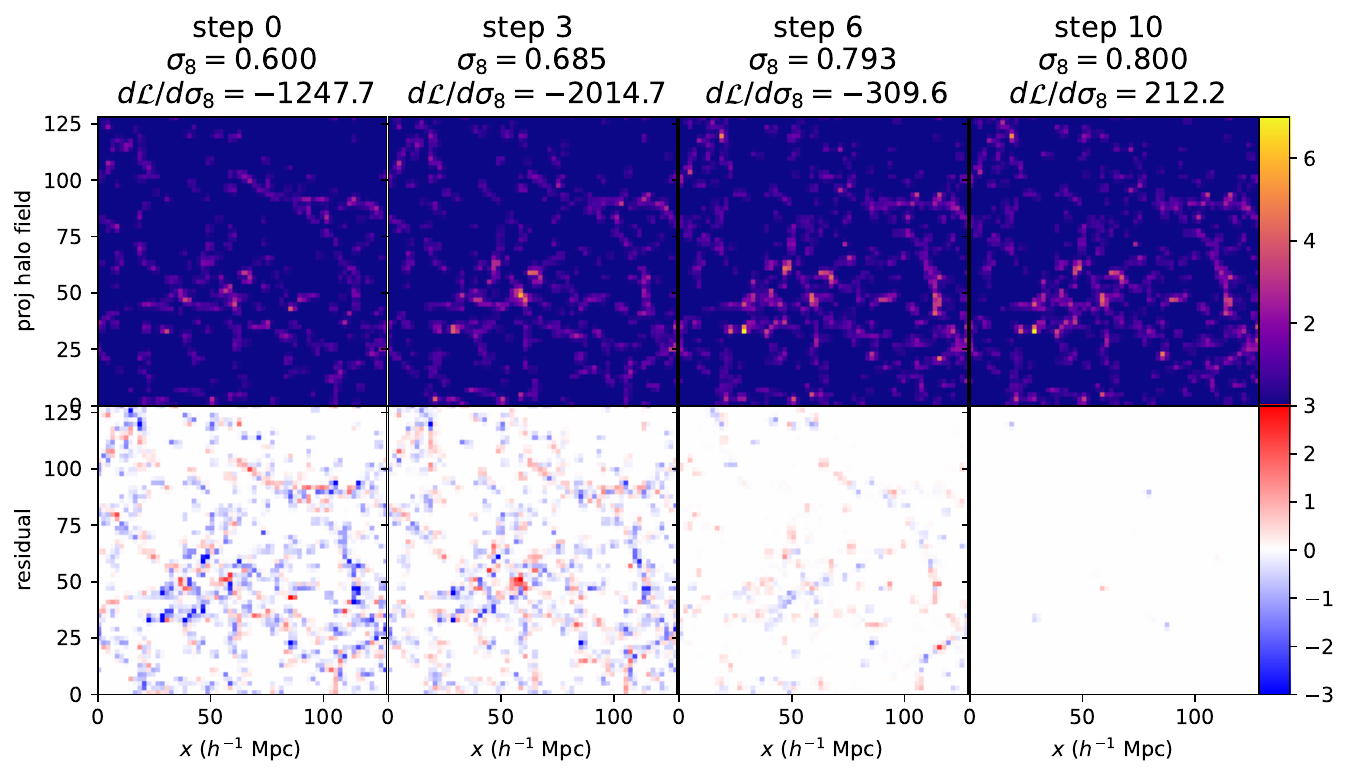}
    \caption{Demonstration of differentiating through frozen assignment for optimization of a simple halo based field-level loss function at fixed initial phases. Each column corresponds to an optimization step with varying $\sigma_8$. The top panels show the projected halo field; bottom panels show residuals relative to the target. The reported $d \mathcal{L}/d\sigma_8$ values trace the sign and magnitude of the gradient as the optimizer converges toward the simulated true value ($\sigma_8 = 0.8$).}
    \label{fig:simple_optimizer}
\end{figure*}
\begin{figure}
    \centering
    \includegraphics[width=0.95\linewidth]{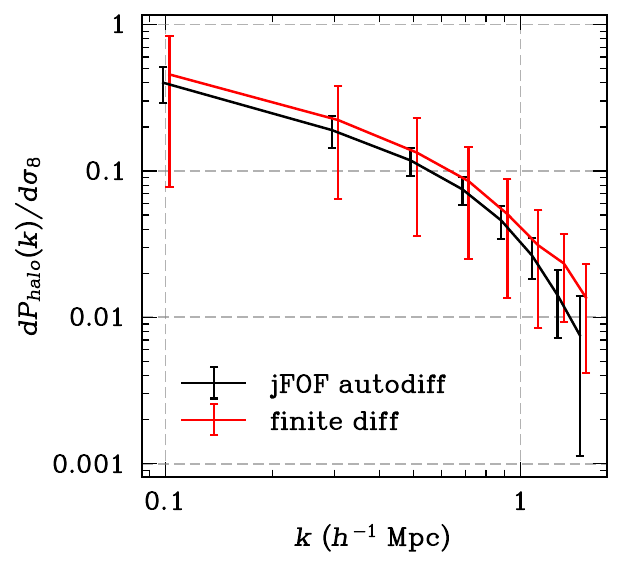}
    \caption{Comparison of the derivative of the halo power spectrum with respect to $\sigma_8$, computed via \texttt{jFoF} autodiff (black) and finite-difference estimation (red). The curves show ensemble averages over 100 realizations. The strong agreement and reduced noise in the autodiff results confirm that frozen-assignment differentiation provides stable and physically meaningful gradients.}
    \label{fig:dpdsigma8}
\end{figure}

A central feature of these methods is the separation between the forward and backward passes. The forward pass executes the standard clustering algorithm exactly—using either the $k$-d  tree or linked-cell neighbor search—and produces halo catalogs identical to those generated by conventional FoF codes. The backward pass, by contrast, focuses solely on estimating gradients. We do not attempt to differentiate through discrete data structures such as the $k$-d  tree or cell assignments. Instead, these quantities are fixed from the forward pass, while particle–halo memberships are either held constant (\emph{frozen assignment}) or allowed to vary under controlled perturbations (\emph{topological optimization}).

Since no analytic gradients exist for these operations, our approach is necessarily empirical. We evaluate whether the estimated gradients capture meaningful dependencies and whether they can accelerate convergence in realistic inference settings. This assessment is inherently problem-specific and requires further experimentation, so here we present proof-of-concept demonstrations illustrating potential techniques rather than a general prescription.

Finally, we note that differentiating full simulation fields through automatic differentiation frameworks such as \texttt{jax} can be substantially more memory intensive than computing scalar likelihoods. This arises from the way \texttt{jax} implements reverse-mode differentiation via vector–Jacobian products (VJPs). During the forward pass, \texttt{jax} traces every intermediate array operation to build a computational graph, storing all activations needed to reconstruct partial derivatives during the backward pass. For high-dimensional fields—such as particle positions or density grids—this graph can easily reach tens or hundreds of gigabytes, as every step in the neighbor search or particle update must be retained in memory until the gradient is evaluated. In contrast, scalar likelihoods or low-dimensional summary statistics can often be differentiated using a single VJP, which contracts the output sensitivities with the relevant Jacobian blocks on the fly, greatly reducing storage overhead. 


\subsection{Frozen Assignment}
\label{subsec:FA}

The most straightforward approach to derivative propagation is to simply leave halo membership fixed during back-propagation. This means that the only valid derivatives are those that are direct functions of gradients of particle positions. The most natural quantity to estimate with this sort of formalism is halo position, i.e.

\begin{equation}
    \vec{X}^h = \frac{1}{N}\sum_{i=0}^N \vec{x}_i \longrightarrow \frac{d\vec{X{^h}}}{d \theta} = \frac{1}{N}\sum_{i=0}^N \frac{d \vec{x}_i}{d \theta},
\end{equation}
where $\vec{X}$ is the position of a halo with $N$ particles located at position $\vec{x}$ and $\theta$ are model parameters to optimize. We note that, in the absence of stochasticity in the optimizer/sampler, this form of gradient calculation is highly restrictive in terms of parameter space accessible. Even so, we find that derivatives of halo position can provide useful information even without mass/merger property propagation.

As a simple demonstration, we perform field level inference of a cosmological parameter at fixed initial phases. For this test, we use \texttt{JaxPM} to evolve an initial density field of boxlength of $128$ $h^{-1}$ Mpc and particle resolution $64^3$ to redshift $z=0$ with $\sigma_8=0.8$. We then use our $k$-d tree halo finder with $b=0.2$ to extract a halo catalog. We perform a mass cut of $N=5$ particles, and then paint our field to a $64$ sidelength grid. 

We treat this halo catalog as baseline observed data and try to fit cosmological parameter, $\sigma_8$, at fixed initial phases starting at $\sigma_8=0.6$. While a straightforward task for a binary search, we use gradient information from our model to inform parameter updates via gradient descent. We find even with the very limited information from bare frozen assignment we can still converge to a reasonable solution within a few function evaluations. 

It is worth further exploring the information content available in the frozen assignment scheme. One way to evaluate it to see how well they can capture summary statistics, like the power spectra. It is worth noting that evaluating Jacobians of summary statistics is generally quite expensive within reverse autodiff frameworks since they are not naturally expressible in terms of vector products. 

We can calculate the expectation value of the Jacobian, $dP(k)/d\sigma_8$ over initial conditions and compare it to the expectation value from finite difference. We show an average over 100 realizations in Figure \ref{fig:dpdsigma8}. We note that, due to the discrete nature of the halo field, finite differences are extremely noisy compared to the Jacobian, but we find the same overall trend. While the trend is very similar between the derivative calculated from finite differences and that from autodiff, there is a systematic offset. This could be explained by the change in overall number density that can be captured by the finite difference formalism that can't be captured in the frozen assignment scheme.

\begin{figure}
    \centering
    \includegraphics[width=1.0\linewidth]{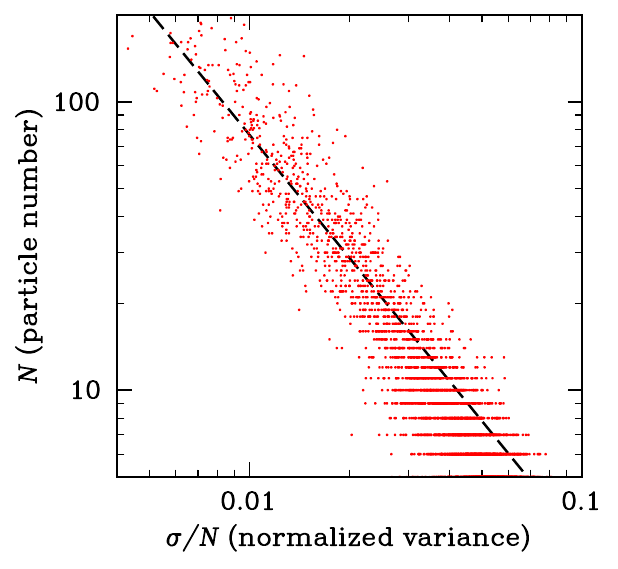}
    \caption{Relationship between the normalized particle variance and the particle number in our simulation. The inverse trend supports using $\sigma$ as a differentiable proxy (“decoration”) for halo mass in the decorated frozen-assignment scheme.}
    \label{fig:decoration}
    \centering
    \includegraphics[width=0.95\linewidth]{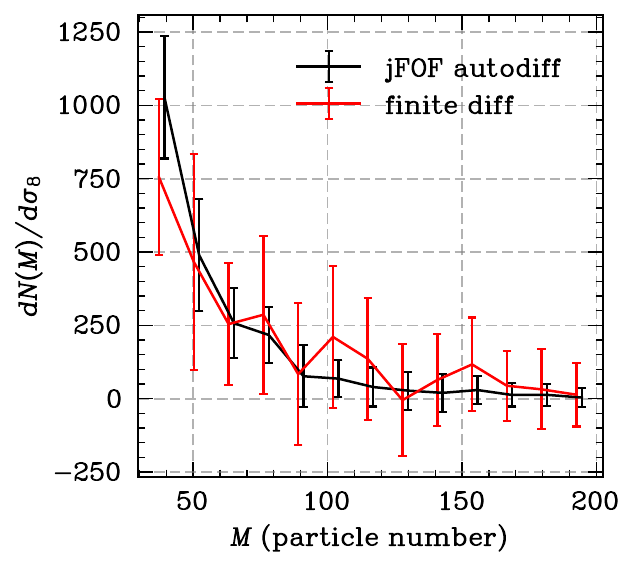}
    \caption{Comparison of the derivative of the halo mass function with respect to $\sigma_8$ from \texttt{jFoF} automatic differentiation with our decorated frozen assignment scheme (black) and those calculated from a finite difference scheme (red). The mean and variance are shown across 100 realizations. Both approaches show consistent trends across the halo mass range, validating the surrogate mass-derivative approach.}
    \label{fig:dndsigma} 
\end{figure}

\subsubsection{Decorated Frozen Assignment}

We can extend this approach beyond halo position to halo mass by looking for suitable proxies for halo mass which are direct analytical functions of the particle positions, $x$. These properties can be calculated on the forward pass and held as a halo properties to vary, which we call ``halo decorations,''($\mathcal{D})$. Schematically we can write derivatives of any halo property ($P$) in terms of this decoration as;
\begin{equation}
  \frac{d P^h}{d\theta} \propto \frac{d P^h}{d \mathcal{D}} \frac{d \mathcal{D}}{d \vec{x}_i} \frac{d\vec{x}_i}{d\theta}.
\end{equation}
For the purpose of this work, we focus on a single decoration, particle standard deviation ($\sigma$) as surrogate for halo mass ($M$). We show the relationship between these properties in the simulation described in Sec \ref{subsec:FA} in Figure \ref{fig:decoration}. Note that we show the standard deviation normalized by particle to correct for the overall increased spread of halos as their particle number increases. The clear inverse trend shown motivates using $\sigma$ as an inverse decoration; if ${d \sigma}/{d \theta}$ is positive then the halo is spreading out and its mass is likely to decrease, and conversely if negative it is further collapsing at likely to get more massive.

With a decoration for halo mass in hand, we can calculate derivatives of downstream halo mass-based quantities in terms of underlying parameters and initial condition fields. One of the simplest to calculate is the halo mass function, a histogram of halo masses commonly used in cosmology and galaxy evolution studies. We apply this approach to the same simulation set discussed in Sec. \ref{subsec:FA}. We use a ``soft" histogram function for back-propagation to ensure derivative information flows between bins.

Like in Sec. \ref{subsec:FA}, we can compare our automatically calculated derivatives with finite differences which we show in Figure \ref{fig:dndsigma}. Again, we find our automatic derivatives offer lower variance than those calculated with finite difference, but recover the same overall trend.

\subsection{Topological Optimization with REINFORCE}
\label{sec:topo}

One general solution to the problem of differentiating through stochastic, non-differentiable computations is the \textit{REINFORCE estimator}, also known as the score-function or likelihood-ratio gradient estimator \citep{williams1992simple}. It belongs to a broad class of methods for estimating gradients of expectations with respect to parameters of a random process, complementing so-called pathwise or reparameterization gradients used in continuous latent-variable models. Whereas reparameterization-based methods rely on expressing a random variable as a deterministic function of parameters and auxiliary noise (allowing gradients to flow through the sampling path), REINFORCE applies to the much more general case where the sampling mechanism itself is discrete or opaque to differentiation.

This stochastic formulation also situates our method within the broader field of \emph{topological optimization}, where the objective depends not only on continuous variables but also on the underlying connectivity or topology of a system \citep{sigmund2013topology}. In classical topology optimization problems, such as structural design \citep{bendsoe1999material,allaire2001some} or graph rewiring \citep{chan2016optimizing}, one seeks to optimize both geometry and discrete structure simultaneously, but the discrete component traditionally breaks differentiability. By introducing stochastic connectivity and optimizing its expected objective via REINFORCE, we obtain a differentiable relaxation of this otherwise combinatorial problem. In the context of halo inference, this allows the optimizer to explore smooth gradients across changes in the discrete topology of the halo graph: halos can appear, merge, or fragment continuously in expectation.

If \(y\) denotes a discrete random outcome (e.g.\ a sampled halo graph) drawn from a parameterized distribution \(p_\theta(y)\), then for any loss function \(L(y)\),
\begin{equation}
\nabla_\theta \, \mathbb{E}_{y\sim p_\theta(y)}[L(y)] 
= \mathbb{E}_{y\sim p_\theta(y)}[L(y)\,\nabla_\theta \log p_\theta(y)].
\end{equation}
This identity allows gradients to propagate through stochastic sampling without requiring differentiability of the discrete step itself. 
In practice, we approximate the expectation with \(S\) Monte Carlo samples, introducing a baseline \(b\) to reduce variance:
\begin{equation}
\nabla_\theta L \;\approx\; 
\frac{1}{S}\sum_{s=1}^{S} (L_s - b)\,\nabla_\theta \log p_\theta(y_s).
\label{eq:reinforce_sampled}
\end{equation}

We exploit this estimator to make the friends-of-friends (FoF) halo-finding process differentiable in expectation. 
Rather than using a hard linking criterion, we define a stochastic surrogate where each potential edge between nearby particles is modeled as a Bernoulli random variable,
\begin{equation}
y_{ij} \sim \mathrm{Bernoulli}(p_{ij}), 
\qquad 
p_{ij} = \sigma\!\left(\alpha\,\frac{b^2 - d_{ij}^2}{|b^2|}\right),
\label{eq:bernouli}
\end{equation}
with linking probability \(p_{ij}\) that varies smoothly with inter-particle distance \(d_{ij}\). 
This formulation makes the \emph{expected} halo connectivity a continuous function of particle positions. 
Using REINFORCE to estimate gradients through the sampled edges allows the optimization to adjust not only the internal properties of halos, as in frozen-assignment differentiation, but also their \textit{number, topology, and membership}. 

\begin{figure}
    \centering
    \includegraphics[width=0.45\linewidth]{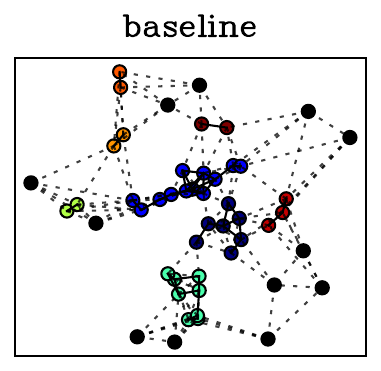}
    \includegraphics[width=0.74\linewidth]{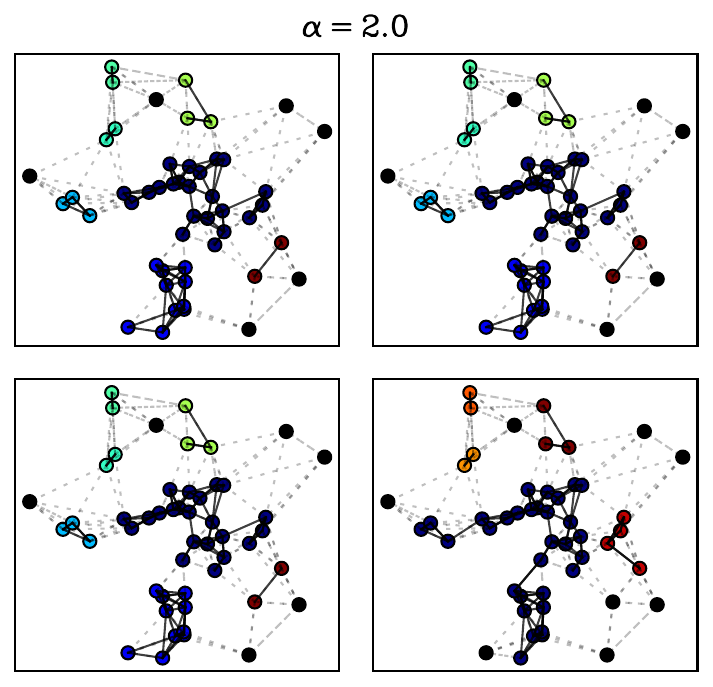}
    \includegraphics[width=0.74\linewidth]{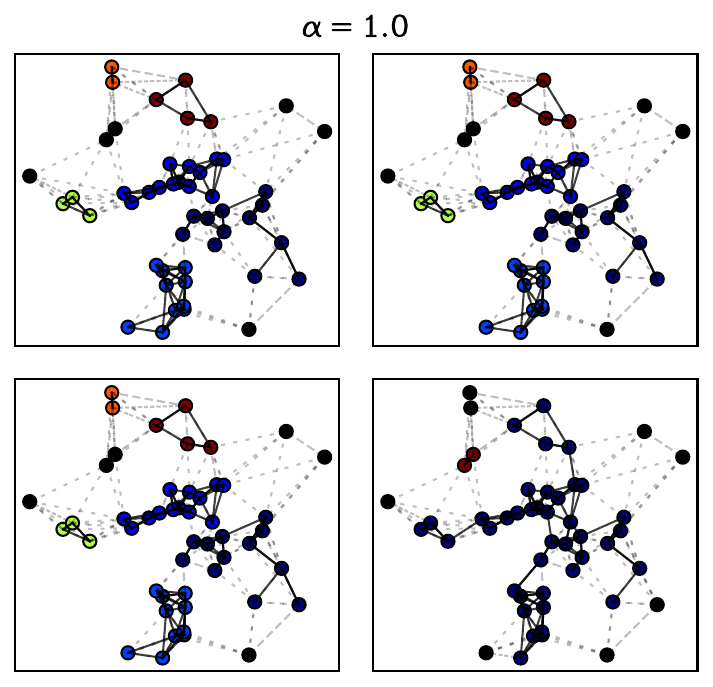}
    \includegraphics[width=0.74\linewidth]{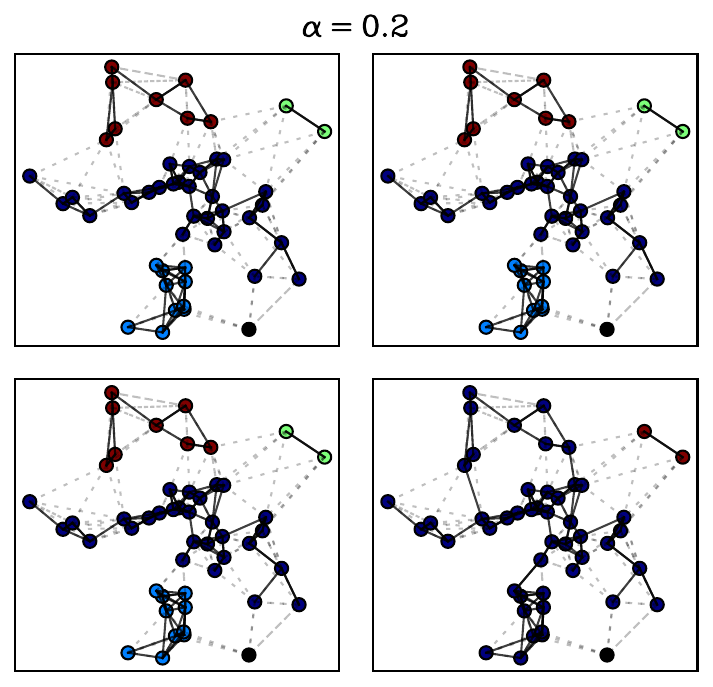}
    \caption{A simple distribution showing the baseline halos (top) and samples varying $\alpha$ from Eq. \ref{eq:bernouli}. High $\alpha$ values correspond to small changes in topology, while lower values push  the distribution closer to random edge assignments.}
    \label{fig:alpha}
\end{figure}

Thus, REINFORCE extends the differentiable regime of structure formation models beyond fixed combinatorial assignments, enabling continuous optimization through discrete merging and fragmentation events. We show some examples of these behaviors in Figure \ref{fig:alpha}, where we vary $\alpha$ and show the various sampled connected edges and halo assignments. Note that as we decrease $\alpha$, we see higher network connectivity (fewer large mass halos) even though the overall number of connections is fairly similar. This is similar to the well known property of ``small-world networks" \citep{watts1998collective} and random graph theory more broadly. Note that we do not rebuild the neighbor list in this implementation, significantly speeding up the algorithm but moderately limiting the set of topological deformations possible. 

\begin{figure}
    \centering
    \includegraphics[width=0.99\linewidth]{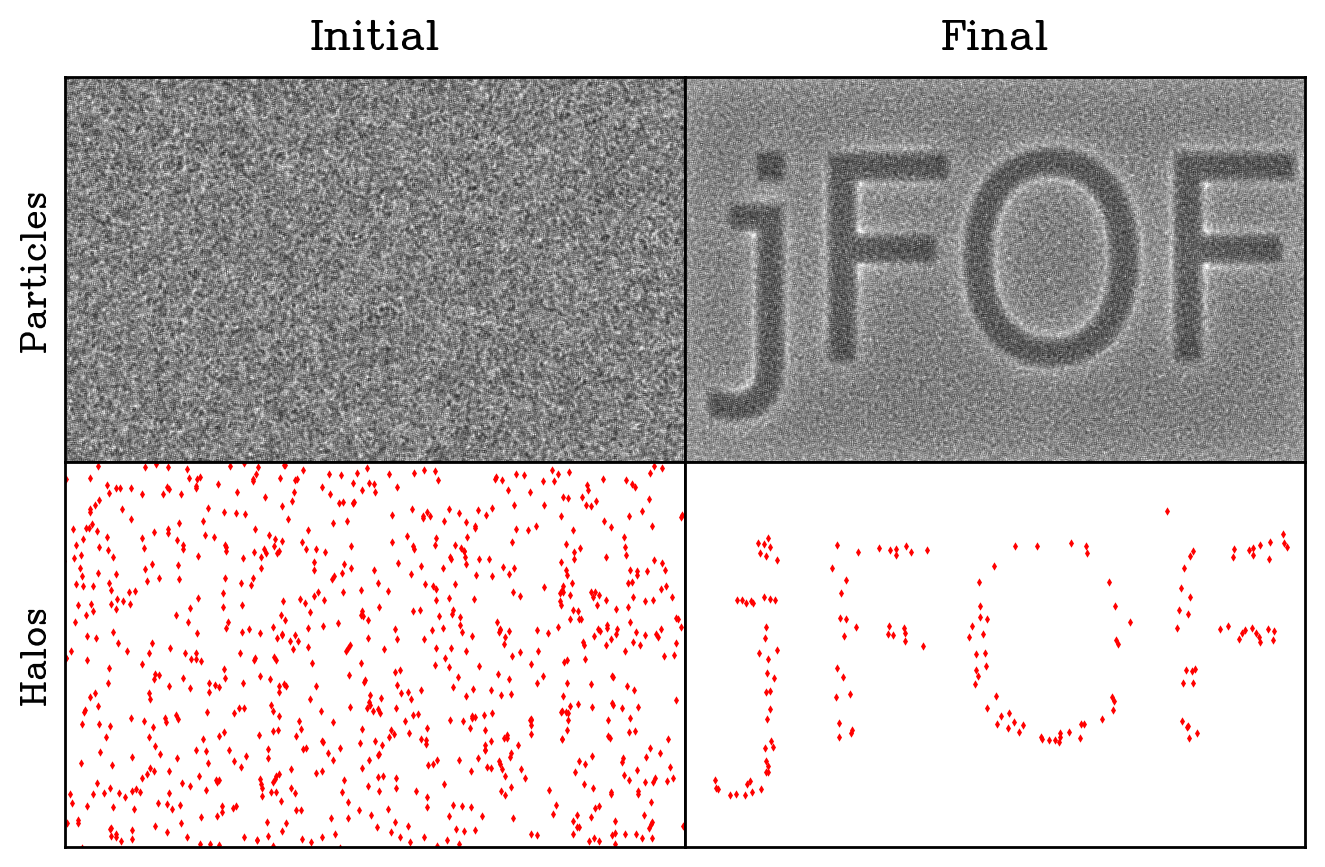}
    \caption{A toy demonstration of using a topological optimization scheme with a field-based loss. Here we penalize halo formation outside of the target image and reward for halo formation within the image. After 300 iterations, halos have reassembled to trace the desired pattern, showcasing \texttt{jFoF}’s ability to modify topology via differentiable probabilistic linking.}
    \label{fig:topo}
\end{figure}
This approach, and REINFORCE more generally, provides an \emph{unbiased} estimate of the true gradient (averaged over many samples, it converges to the correct value) but it can exhibit relatively high variance, since each Monte Carlo realization contributes a noisy estimate of the loss-weighted log-probability term. In practice, variance is mitigated by using multiple samples per step, introducing a control variate or baseline (often the batch-mean loss, as in this work), and normalizing or clipping the resulting advantage. These variance-reduction techniques preserve unbiasedness while making the gradient stable enough for optimization in high-dimensional, stochastic systems like differentiable halo-finding.

Due to the high variance of the gradients, significant care should be taken in likelihood setup and optimization/sampling design. These techniques require more careful tuning than the frozen-assignment schemes discussed in Sec. \ref{subsec:FA}. Convergence properties will depend on the choice of number of samples, $S$, relaxation parameter, $\alpha$, the functional form chosen for $\sigma$, and also on the underlying FoF parameters which drive the linking probability dimension and sparsity.

\subsubsection{Toy Problem}

To illustrate the capabilities of our approach in a controlled setting, we first apply it to a simplified optimization problem designed to minimize a straightforward loss function without introducing confounding physical effects. Starting from a uniform random particle distribution, our goal is to move particles such that they form halos above a specified minimum mass threshold, following a prescribed target spatial distribution, $\Omega$. We specify a loss function of the form;

\begin{equation}
    \mathcal{L} = \sum_{H\notin\Omega} M_H -\sum_{H\in\Omega} M_H,
\end{equation}

where $M_H$ is the mass of the halo, $H$. Although conceptually simple, this task is highly nontrivial within a differentiable optimization framework. The model must not only move particles toward overdense regions to form new halos, but also fragment existing halos that lie outside the target distribution. To explicitly encourage this behavior, we chose a mass weighting in the loss function rather than a pure halo count. This term rewards the redistribution of mass away from halos outside the target region—even if such movement does not immediately lead to halos below the mass cutoff.
\label{subsubsec:dynamic}
\label{sec:fieldlevel}
\begin{figure*}
    \centering
    \includegraphics[width=0.995\linewidth]{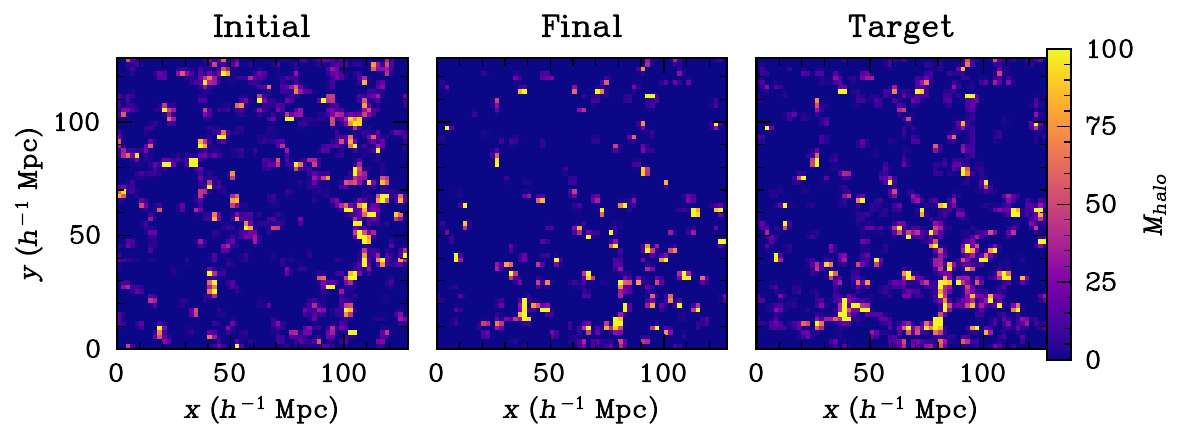}
    \caption{A projection of the three-dimensional dynamic field level halo reconstruction described in Sec. \ref{sec:fieldlevel} using the REINFORCE-based approach. The panels show the projected halo field at initialization, after optimization (500 steps), and for the target field. The final result successfully recovers large-scale structure morphology and halo clustering.}
    \label{fig:3panel_tl}
\end{figure*}

\begin{figure}
    \centering
    \includegraphics[width=0.99\linewidth]{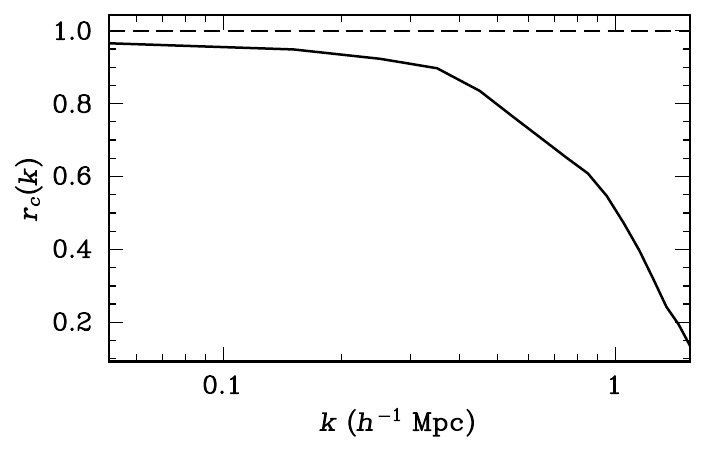}
    \caption{Correlation coefficient of the resulting mass-weighted halo field after ~500 iterations. High correlation at large and intermediate scales demonstrates the effectiveness of the topological loss for field-level inference.}
    \label{fig:rc_tl}
\end{figure}
To evaluate the scalability of the method, we initialize $64^3$ particles on a two-dimensional grid spanning the same extent as the target distribution. At each iteration, we perform halo identification and compute the approximate gradient using $S = 16$ Monte Carlo samples, a relaxation parameter of $\alpha = 0.5$, and a smooth sigmoidal thresholding function $\sigma$. The optimization is run for 300 iterations, and the resulting reconstructed fields are presented in Figure~\ref{fig:topo}. We find essentially all halos outside the target region fragment, while particles are pushed into the target region for additional halo formation.

\subsubsection{Dynamic Halo Field Reconstruction}

We an also use this technique to perform a full field level inference of the halo field from cosmological initial conditions. When combined with a differentiable halo occupancy distribution (i.e. \citep{2024MNRAS.529.2473H}), this would constitute an end-to-end model for galaxy field level inference. There is significant ongoing effort in this direction from the standpoint of bias modeling \citep{Schmidt:2018bkr, Schmidt:2020viy, 2021JCAP...03..058N, Jeffrey:2020xve, Kostic:2022vok,VelRecon,2024PhRvL.133v1006N, 2025arXiv250420130S}, but comparatively little that attempt to model the galaxies beyond a bias model description (see \citet{2018JCAP...10..028M} for a notable exception, and the work of \cite{Parker:2025mtg} who directly trained a CNN from galaxies to initial conditions to perform reconstruction). In this section, we do not attempt to solve these issues completely, but instead demonstrate that it is possible to use a topological loss into a field level inference pipeline.

For this experiment, we adopt the same simulation configuration described in Section~\ref{subsec:FA}. We take as input a representative $z = 0$ mass-weighted halo field, imposing a minimum particle threshold of $N = 5$, add shot noise based on Poisson noise, and treat this as our mock data field.

For our forward model, we run our initial conditions at each step forward with \texttt{jaxpm} to $z=0$ and find halos with the same critera as our mock generation. To compare this theoretical prediction to our mock halo-mass counts, use a simple mean squared error loss on the resulting mass-weighted halo field.

We employ our REINFORCE-based optimizer, averaging over $S = 16$ realizations with a relaxation parameter $\alpha = 5.0$. Optimization is performed using the \texttt{Adam} algorithm with a learning rate of $0.2$. To improve convergence, we apply an annealing schedule to the data smoothing scale: starting with a Gaussian smoothing of $4~h^{-1},\mathrm{Mpc}$ and progressively reducing the smoothing radius every 50 iterations, until reaching an unsmoothed field after 250 iterations. We show results after 500 iterations, corresponding to a total of 8500 halo-find operations (although only requiring 500 tree-construction operations).

The results are shown in Figure~\ref{fig:3panel_tl}, where we display the projected mass-weighted halo field before and after optimization. The reconstructed field exhibits substantial deviation from the random initialization, indicating effective learning of structural features.  This improvement is also reflected quantitatively in the cross-correlation coefficient (Figure~\ref{fig:rc_tl}), which demonstrates that our topological loss enables recovery of information across a broad range of spatial scales.

\section{Conclusions and Discussion}

We have presented \texttt{jFoF}, a GPU-native Friends-of-Friends halo finder built for both high-performance simulation analysis and differentiable modeling. By implementing both k–d tree and linked-cell neighbor search schemes within \texttt{jax}, we demonstrated that large-scale halo identification can be performed entirely on GPUs without costly CPU-device transfers. The resulting system achieves timing performance significantly better than traditional distributed-memory CPU frameworks while maintaining full consistency in halo identification. These results confirm that FoF halo finding can be efficiently reimagined for modern accelerator architectures without sacrificing accuracy or fidelity.

Beyond raw performance, \texttt{jFoF} introduces a novel capacity for gradient propagation through otherwise discrete halo identification procedures. By implementing “frozen assignment” and “topological optimization” modes, we have demonstrated how derivative information (normally inaccessible in discontinuous clustering operations) can be estimated and used in optimization-based inference. These approaches allow gradients of halo positions and surrogate mass proxies to flow back through simulation pipelines, forming the foundation for fully differentiable inference framework for cosmology and/or galaxy evolution. Such functionality opens a new regime where the structure-formation process can be treated as a continuous, optimizable transformation rather than a fixed post-processing step.

Our gradient propagation approach can be plugged into gradient-based Monte Carlo samplers to differentiate directly through the halo finding instead of using bias models to approximate the halo field in the forward model. It is important to stress that, in the context of a sampling algorithm such as Hamiltonian Monte Carlo, noise in the gradient calculation is not necessarily detrimental to convergence and in some cases will improve overall performance \citep{chen2014stochastic}. In the context of samplers with a metropolis-hastings type acceptance step, the most important thing is accuracy in the likelihood calculation and, by extension, the forward model itself. The methods explored in this work for gradient estimation do not affect the exactness in the forward path.

Our results highlight a path toward unified GPU-resident workflows in which simulation, analysis, and inference coexist within a single differentiable computational graph. When coupled with differentiable N-body or hydrodynamic solvers such as \texttt{FlowPM} \citep{2021A&C....3700505M}, \texttt{pmwd} \citep{2022arXiv221109958L}, \texttt{DISCO-DJ} \citep{2024discodj,2025discodj2}, or \texttt{DiffHydro} \citep{2025arXiv250202294H} and differentiable HOD codes \citep{2024MNRAS.529.2473H} or population modeling \citep{2023MNRAS.518..562A,2023MNRAS.521.1741H}, \texttt{jFoF} provides the missing link needed to perform end-to-end optimization directly on structure catalogs. This integration will enable joint constraints of physical parameters or subgrid models through gradient-based likelihood maximization, rather than relying on discrete sampling or empirical calibration. In practice, this could enable significantly faster model tuning and more interpretable connections between cosmological initial conditions and observable structures beyond bias models commonly used.

It should be noted that the convergence properties of the pure topological optimizer for dynamical problems like those in Sec. \ref{subsubsec:dynamic} quite complex and intuition built in the bias model approach (i.e. \citet{Schmidt:2018bkr}) isn't necessarily directly transferable. There are a number of additional numerical factors to tune in both the REINFORCE gradient and the FoF halo finder itself which could affect convergence properties. In a regime where there are few particles near the target halo distribution, it is unlikely that the sampled reinforce loss (i.e. Eq \ref{eq:reinforce_sampled}) will successfully reach the minimum halo mass threshold. In practice, a hybrid approach wherein a bias model formalism or Kaiser-style preconditioning \citep{2025arXiv250420130S} is used at first and then switched to a topological loss later on in the optimization to push to smaller scales in the reconstruction. We leave an exploration of these properties for future work.

In future work, we plan to extend \texttt{jFoF} with adaptive linking schemes, subhalo merger tracking, and fully probabilistic relaxation methods based on Gumbel–Softmax reparameterization. These will enable smoother topological transitions and (hopefully) improved stability in gradient estimation. Combining these features with distributed GPU parallelism and memory-efficient streaming label propagation will further enhance scalability toward simulations exceeding $10^9$ particles. Ultimately, \texttt{jFoF} aims to serve as a key tool for GPU-based large-scale structure analysis; uniting high-performance computing and machine learning principles to build the next generation of cosmological inference frameworks.

\section*{Acknowledgements}
We thank Benjamin Dodge and Francois Lanusse for very useful discussions and suggestions. We thank the Erwin Schrödinger International Institute for Mathematics and Physics (ESI) where this project was partially completed.

\section*{Data Availability}

No new data were generated or analysed in support of this research. All code related to this work is available at \url{https://github.com/bhorowitz/jFOF/tree/main}.

\bibliographystyle{mnras}
\bibliography{example} 






\bsp	
\label{lastpage}
\end{document}